%% file: main.tex
\documentclass{article}
\usepackage{spconf,amsmath,graphicx}
\usepackage{mathtools}
\usepackage{empheq}
\usepackage{bbold}
\usepackage{color}
\usepackage[hidelinks]{hyperref}
\usepackage{booktabs}
\usepackage{tabularx}
\usepackage{units}
\usepackage{pgfplots}
\pgfplotsset{width=7cm,compat=1.8}
\usepackage{cite} % for [1-3] instead of [1,2,3]
\usepackage{dblfloatfix}

\newcommand*{\bPhi}[1]{\boldsymbol{\Phi}_{\mathbf{#1},l}}
\newcommand*{\bhPhi}[1]{\widehat{\boldsymbol{\Phi}}_{\mathbf{#1},l}}
\newcommand*{\bgamma}[1]{\boldsymbol{\gamma}_{\mathbf{#1},l}}
\newcommand*{\bhgamma}[1]{\widehat{\boldsymbol{\gamma}}_{\mathbf{#1},l}}
\newcommand*{\hxi}{\widehat{\xi}_l}
\newcommand*{\mbf}[1]{\mathbf{f}^{l}_{\mathbf{#1}}}
\newcommand*{\f}[1]{f^{l}_{#1}}
\newcommand*{\bh}[1]{\mathbf{h}_{\mathbf{#1},l}}
\newcommand*{\bhh}[1]{\widehat{\mathbf{h}}_{\mathbf{#1},l}}
\newcommand*{\bhH}[1]{\widehat{\mathbf{H}}_{\mathbf{#1},l}}
\newcommand*{\tran}{^{\mkern-1.5mu\mathsf{T}}}
\newcommand*{\conj}{^{\mathsf{*}}}
\newcommand*{\hermconj}{^{\mathsf{H}}}

\title{Deep Multi-Frame MVDR Filtering for\\
	Single-Microphone Speech Enhancement}
\name{Marvin Tammen, Simon Doclo
	\thanks{This work was funded by the Deutsche Forschungsgemeinschaft (DFG, German Research Foundation) – Project ID 390895286 – EXC 2177/1.}}
\address{Department of Medical Physics and Acoustics and Cluster of Excellence Hearing4all\\
	University of Oldenburg, Germany\\$\lbrace$marvin.tammen, simon.doclo$\rbrace$@uni-oldenburg.de}
\begin{document}
\ninept
\maketitle
\begin{abstract}
	Multi-frame algorithms for single-microphone speech enhancement, e.g., the multi-frame minimum variance distortionless response (MFMVDR) filter, are able to exploit speech correlation across adjacent time frames in the short-time Fourier transform (STFT) domain.
	Provided that accurate estimates of the required speech interframe correlation vector and the noise correlation matrix are available, it has been shown that the MFMVDR filter yields a substantial noise reduction while hardly introducing any speech distortion.
	Aiming at merging the speech enhancement potential of the MFMVDR filter and the estimation capability of temporal convolutional networks (TCNs), in this paper we propose to embed the MFMVDR filter within a deep learning framework.
	The TCNs are trained to map the noisy speech STFT coefficients to the required quantities by minimizing the scale-invariant signal-to-distortion ratio loss function at the MFMVDR filter output.
	Experimental results show that the proposed deep MFMVDR filter achieves a competitive speech enhancement performance on the Deep Noise Suppression Challenge dataset.
	In particular, the results show that estimating the parameters of an MFMVDR filter yields a higher performance in terms of PESQ and STOI than directly estimating the multi-frame filter or single-frame masks and than Conv-TasNet.
\end{abstract}
\section{Introduction}
In many hands-free speech communication systems such as hearing aids, mobile phones and smart speakers, ambient noise may degrade the speech quality and intelligibility of the recorded microphone signals. 
To alleviate this issue, several single- and multi-microphone speech enhancement algorithms have been proposed ~\cite{vary_digital_2006,hendriks_dft-domain_2013,benesty_speech_2011,doclo_multichannel_2015,vincent_audio_2018}. 
Single-microphone speech enhancement algorithms typically 1) transform the noisy time-domain signal to a domain that is better suited for speech enhancement, e.g., the short-time Fourier transform (STFT) domain, 2) apply a (real- or complex-valued) gain/mask to the transform-domain coefficients to obtain an estimate of the clean speech, and 3) transform the modified coefficients back to the time-domain.
For such single-frame algorithms, many traditional model-based approaches~\cite{vary_digital_2006,hendriks_dft-domain_2013,ephraim_speech_1984,gerkmann_phase_2015} as well as supervised learning-based approaches~\cite{xu_regression_2015,wang_training_2014,kolbaek_speech_2017,luo_conv-tasnet_2019,tan_gated_2019} have been proposed.
A disadvantage of single-frame algorithms is that attenuation of the noise component may be accompanied by some distortion of the speech component in the enhanced signal. 

In contrast to single-frame algorithms, multi-frame algorithms have been proposed which apply a complex-valued filter to the noisy speech STFT coefficients~\cite{benesty_speech_2011}.
Also for multi-frame algorithms both model-based approaches such as the multi-frame minimum variance distortionless response (MFMVDR) filter~\cite{huang_multi-frame_2012,schasse_estimation_2014,fischer_sensitivity_2017,fischer_subspace-based_2020} as well as supervised learning-based approaches~\cite{mack_deep_2019,xu_neural_2020} have been proposed.
The MFMVDR filter has been derived to explicitly take speech correlations across adjacent time frames into account and requires an estimate of the noise correlation matrix and the so-called speech interframe correlation (IFC) vector in each time-frequency bin.
Although it has been shown that the MFMVDR filter can yield a good noise reduction performance and little speech distortion~\cite{huang_multi-frame_2012,fischer_sensitivity_2017}, its performance is very sensitive to estimation errors of the required quantities, in particular the speech IFC vector~\cite{fischer_sensitivity_2017}. 

To estimate the speech IFC vector from the noisy speech STFT coefficients, several model-based approaches have been proposed.
In \cite{schasse_estimation_2014} a maximum likelihood approach has been derived, assuming that the speech and noise IFC vectors follow multi-variate Gaussian distributions.
This approach requires an estimate of the a-priori signal-to-noise ratio (SNR), which can be estimated, e.g., using the decision-directed approach~\cite{ephraim_speech_1984} or using a supervised learning-based approach~\cite{tammen_dnn-based_2020}.
In \cite{fischer_subspace-based_2020} a subspace estimator has been proposed, which is based on a low-rank speech model.
However, simulation results have shown that estimating the required quantities from the noisy speech STFT coefficients using these model-based approaches typically results in a large performance degradation compared to the oracle MFMVDR filter.

\begin{figure*}
	\centering
	\def\svgwidth{\linewidth}
	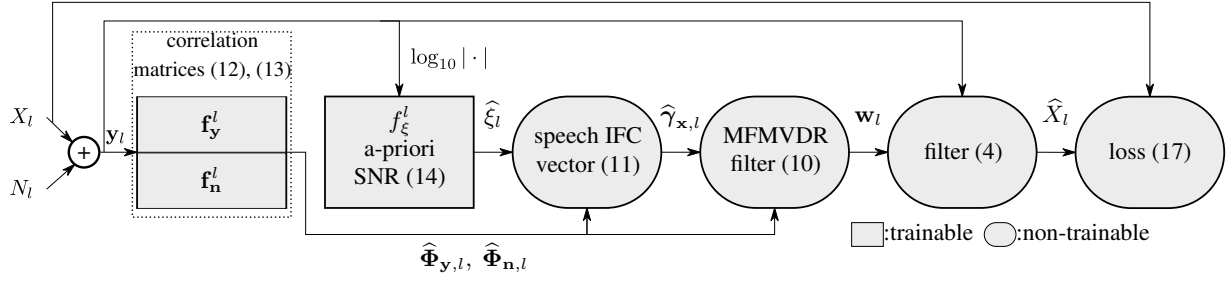
	\caption{Block diagram of the training process of the proposed deep MFMVDR filter. 
		The speech enhancement-related loss function is used to update the weights of the temporal convolutional networks estimating the noisy speech and noise correlation matrices $\bPhi{y}$ and $\bPhi{n}$ as well as the a-priori SNR $\xi_l$.}
	\label{fig:block diagram}
\end{figure*}

In this paper we propose to embed the MFMVDR filter within a deep learning framework as shown in Fig. \ref{fig:block diagram}.
More in particular, we propose to train temporal convolutional networks~\cite{bai_empirical_2018,luo_conv-tasnet_2019} to map the noisy speech STFT coefficients to the required quantities, i.e., the noise correlation matrix and the a-priori SNR, by minimizing the scale-invariant signal-to-distortion ratio loss function~\cite{roux_sdr_2019} at the MFMVDR filter output.
Experimental results using the INTERSPEECH 2020 Deep Noise Suppression (DNS) Challenge dataset~\cite{reddy_interspeech_2020} show that the proposed deep MFMVDR filter outperforms complex-valued masking as well as directly estimating the multi-frame filter without exploiting the MFMVDR structure and Conv-TasNet~\cite{luo_conv-tasnet_2019}.

\section{Signal Model}
We consider an acoustic scenario with a single microphone recording one speech source and additive ambient noise. 
In the STFT-domain, the noisy microphone signal is given by
\begin{equation}
	\label{eq:signal model}
	Y_{k,l} = X_{k,l} + N_{k,l},
\end{equation}
where $Y_{k,l}$, $X_{k,l}$, and $N_{k,l}$ denote the noisy speech component, the speech component, and the noise component, respectively, at the $k$-th frequency bin and the $l$-th time frame. 
Since all frequency bins are assumed to be independent, the index $k$ will be omitted in the remainder of this paper.

In single-frame speech enhancement algorithms~\cite{vary_digital_2006,hendriks_dft-domain_2013,ephraim_speech_1984,gerkmann_phase_2015,wang_training_2014,kolbaek_speech_2017,luo_conv-tasnet_2019,tan_gated_2019}, the speech component $X_l$ is typically estimated by applying a (real- or complex-valued) gain/mask $W_l$ to the noisy speech STFT coefficients, i.e.,
\begin{equation}
	\label{eq:masking}
	\widehat{X}_{l} = W_l Y_l.
\end{equation}
In multi-frame speech enhancement algorithms~\cite{benesty_speech_2011}, the $N$-dimensional noisy speech vector $\mathbf{y}_l$ is defined as
\begin{equation}
	\label{eq:multi-frame vector}
	\mathbf{y}_l = \left[ Y_l, \ Y_{l-1}, \ \dots, \ Y_{l-N+1} \right]\tran,
\end{equation}
where $\circ\tran$ denotes the transpose operator.
Using \eqref{eq:signal model}, the noisy speech vector $\mathbf{y}_l$ can be written as $\mathbf{y}_{l} = \mathbf{x}_{l} + \mathbf{n}_{l}$, where the speech vector $\mathbf{x}_{l}$ and the noise vector $\mathbf{n}_{l}$ are defined similarly as $\mathbf{y}_l$ in \eqref{eq:multi-frame vector}.
The speech component $X_l$ is then estimated by applying a (complex-valued) finite impulse response filter $\mathbf{w}_l$ with $N$ taps to the noisy speech STFT coefficients, i.e.,
\begin{align}
	\label{eq:speech estimate vector}
	\widehat{X}_{l} &= \mathbf{w}_l\hermconj \mathbf{y}_{l},
\end{align}
where $\circ\hermconj$ denotes the Hermitian operator.

Assuming that the speech and noise components are uncorrelated, the $N \times N$-dimensional noisy speech correlation matrix $\boldsymbol{\Phi}_{\mathbf{y},l} = \mathcal{E} \left\lbrace\mathbf{y}_{l} \mathbf{y}\hermconj_{l}\right\rbrace$, with $\mathcal{E}\{\circ\}$ the expectation operator, can be written as
\begin{equation}
	\boldsymbol{\Phi}_{\mathbf{y},l} = \boldsymbol{\Phi}_{\mathbf{x},l} + \boldsymbol{\Phi}_{\mathbf{n},l},
	\label{eq:mic correlation matrix}
\end{equation}
with the speech and noise correlation matrices $\boldsymbol{\Phi}_{\mathbf{x},l} = \mathcal{E} \left\lbrace\mathbf{x}_{l} \mathbf{x}\hermconj_{l}\right\rbrace$ and $\boldsymbol{\Phi}_{\mathbf{n},l} = \mathcal{E} \left\lbrace\mathbf{n}_{l} \mathbf{n}\hermconj_{l}\right\rbrace$.
In~\cite{huang_multi-frame_2012}, it has been proposed to exploit the speech correlation across adjacent time frames by decomposing the speech vector into a temporally correlated and a temporally uncorrelated part, i.e.,
\begin{equation}
	\mathbf{x}_{l} = \underbrace{\boldsymbol{\gamma}_{\mathbf{x},l} X_{l}}_{\mathclap{\textnormal{correlated}}} \quad + \quad \underbrace{\mathbf{x'}_{l}}_{\mathclap{\textnormal{uncorrelated}}},
	\label{eq:correlated uncorrelated}
\end{equation}
where the (highly time-varying) speech IFC vector $\boldsymbol{\gamma}_{\mathbf{x},l}$ describes the correlation between the current and previous time frames w.r.t. the speech STFT coefficient $X_{l}$, i.e.,
\begin{equation}
	\label{eq:gamma_x}
	\boldsymbol{\gamma}_{\mathbf{x},l} = \frac{\mathcal{E}\left\lbrace \mathbf{x}_{l} X\conj_{l} \right\rbrace}{\mathcal{E}\left\lbrace \left| X_{l} \right|^2 \right\rbrace}
	= \frac{\boldsymbol{\Phi}_{\mathbf{x},l} \mathbf{e}}{\mathbf{e}\tran \boldsymbol{\Phi}_{\mathbf{x},l} \mathbf{e}},
\end{equation}
where $\circ\conj$ denotes the complex-conjugate operator, $\mathbf{e} = [1,\ 0,\ \dots,\ 0]\tran$ is an $N$-dimensional selection vector, and $\mathbf{e}\tran \boldsymbol{\Phi}_{\mathbf{x},l} \mathbf{e} = \mathcal{E} \left\lbrace \left| X_{l} \right|^2 \right\rbrace = \phi_{X,l}$ corresponds to the speech power spectral density (PSD).
Using \eqref{eq:mic correlation matrix} and \eqref{eq:gamma_x}, the speech IFC vector $\boldsymbol{\gamma}_{\mathbf{x},l}$ can be written as
\begin{empheq}[box=\fbox]{align}
	\label{eq:speech IFC vector}
	\boldsymbol{\gamma}_{\mathbf{x},l} = \frac{1 + \xi_l}{\xi_l} \frac{\bPhi{y} \mathbf{e}}{\mathbf{e}\tran \bPhi{y} \mathbf{e}}
	- \frac{1}{\xi_l} \underbrace{\frac{\bPhi{n} \mathbf{e}}{\mathbf{e}\tran \bPhi{n} \mathbf{e}}}_{\bgamma{n}}
\end{empheq}
where $\xi_{l} = \frac{\phi_{X,l}}{\phi_{N,l}}$ denotes the a-priori SNR, with $\phi_{N,l} = \mathcal{E} \left\lbrace \left| N_{l} \right|^2 \right\rbrace = \mathbf{e}\tran \boldsymbol{\Phi}_{\mathbf{n},l} \mathbf{e}$ the noise PSD, and $\bgamma{n}$ denotes the noise IFC vector.

\section{Deep Multi-Frame MVDR Filter}
In~\cite{huang_multi-frame_2012,benesty_speech_2011} the MFMVDR filter for single-microphone speech enhancement has been proposed, which aims at minimizing the output noise PSD while not distorting the correlated speech component $\boldsymbol{\gamma}_{\mathbf{x},l} X_{l}$, i.e.,
\begin{empheq}{align}
	\min_{\mathbf{w}_{l} \, \in \, \mathbb{C}^N} \quad \mathbf{w}\hermconj_{l} \boldsymbol{\Phi}_{\mathbf{n},l} \mathbf{w}_{l}, \quad
	\textnormal{s.t. } \ \mathbf{w}\hermconj_{l} \boldsymbol{\gamma}_{\mathbf{x},l} = 1.
\end{empheq}
Solving this constrained optimization problem yields the MFMVDR filter vector:
\begin{empheq}[box=\fbox]{align}
	\mathbf{w}_{\textnormal{MFMVDR},l} = \frac{\boldsymbol{\Phi}^{-1}_{\mathbf{n},l} \boldsymbol{\gamma}_{\mathbf{x},l}}{\boldsymbol{\gamma}\hermconj_{\mathbf{x},l} \boldsymbol{\Phi}^{-1}_{\mathbf{n},l} \boldsymbol{\gamma}_{\mathbf{x},l}}
	\label{eq:MFMVDR}
\end{empheq}

To implement the MFMVDR filter, estimates of the noise correlation matrix $\bPhi{n}$ as well as the speech IFC vector $\bgamma{x}$ are required. 
In \cite{schasse_estimation_2014,fischer_sensitivity_2017} it has been shown that the speech enhancement performance of the MFMVDR filter strongly depends on how well these quantities can be estimated from the noisy speech STFT coefficients.
Previously proposed model-based approaches~\cite{schasse_estimation_2014,fischer_subspace-based_2020} simply use an estimate of the noisy speech correlation matrix $\bPhi{y}$ instead of the noise correlation matrix $\bPhi{n}$ (leading to the multi-frame minimum \emph{power} distortionless response filter) and estimate the speech IFC vector $\bgamma{x}$ based on \eqref{eq:speech IFC vector} by using the decision-directed approach~\cite{ephraim_speech_1984} to estimate the a-priori SNR $\xi_l$ and assuming the noise IFC vector $\bgamma{n}$ to be constant for all time-frequency points.

In this paper we propose to estimate the required quantities for the MFMVDR filter in \eqref{eq:MFMVDR} from the noisy speech STFT coefficients using a supervised learning-based approach by minimizing a speech enhancement-related loss function at the MFMVDR filter output.

\subsection{Speech IFC Vector}
Due to its highly time-varying nature, the speech IFC vector $\bgamma{x}$ is difficult to estimate accurately.
Since preliminary experiments have shown that using \eqref{eq:speech IFC vector} instead of directly estimating $\bgamma{x}$ with a DNN yields a higher speech enhancement performance, we propose to estimate $\bgamma{x}$ using the estimated correlation matrices $\bhPhi{y}$ and $\bhPhi{n}$ as well as the estimated a-priori SNR $\hxi$, i.e.,
\begin{empheq}[box=\fbox]{align}
	\label{eq:speech IFC vector estimate}
	\bhgamma{x} = \frac{1 + \hxi}{\hxi} \frac{\bhPhi{y} \mathbf{e}}{\mathbf{e}\tran \bhPhi{y} \mathbf{e}}
	- \frac{1}{\hxi} \frac{\bhPhi{n} \mathbf{e}}{\mathbf{e}\tran \bhPhi{n} \mathbf{e}}
\end{empheq}

\subsection{Correlation Matrices}
\label{sec:correlation matrices}
Since the $N \times N$-dimensional correlation matrices $\bPhi{y}$ and $\bPhi{n}$ can be assumed to be Hermitian positive semidefinite (PSD), each matrix consists of a total of $N^2$ real-valued coefficients, denoted by $\bh{y}$ and $\bh{n}$.
As illustrated in Fig. \ref{fig:block diagram}, we propose to estimate these coefficients using separate DNNs $\mbf{y}$ and $\mbf{n}$ in state $l$, i.e.,
\begin{gather}
	\begin{aligned}
		\label{eq:correlation matrix states}
		\bhh{y} &= \mbf{y} \left( \mathbf{y}_{c,l} \right)\\
		\bhh{n} &= \mbf{n} \left( \mathbf{y}_{c,l} \right),
	\end{aligned}
\end{gather}
where $\widehat{\circ}$ denotes an estimate of $\circ$, and $\mathbf{y}_{c,l} =  \left[ \Re Y_l, \Im Y_l \right]\tran$ denotes the vector of the real and imaginary parts of the noisy speech STFT coefficient $Y_l$.
Since the coefficients $\bh{y}$ and $\bh{n}$ are unbounded, a linear activation is used for $\mbf{y}$ and $\mbf{n}$.
The Hermitian PSD correlation matrix estimates are obtained as
\begin{empheq}[box=\fbox]{gather}
	\begin{aligned}
		\label{eq:correlation matrices}
		\bhPhi{y} &= \bhH{y} \bhH{y}\hermconj, \quad \bhH{y} \hspace{-10pt}&&= \textnormal{Hermitian} \left\lbrace \bhh{y} \right\rbrace\\
		\bhPhi{n} &= \bhH{n} \bhH{n}\hermconj, \quad \bhH{n} \hspace{-10pt}&&= \textnormal{Hermitian} \left\lbrace \bhh{n} \right\rbrace
	\end{aligned}
\end{empheq}
where $\textnormal{Hermitian} \left\lbrace \mathbf{h} \right\rbrace$ assembles an $N \times N$-dimensional Hermitian matrix from the (real-valued) $N^2$-dimensional vector $\mathbf{h}$, and the matrix multiplication ensures that the correlation matrix estimates are Hermitian PSD.
It should be noted that no correlation matrix labels are used in the training process --- instead, the DNNs are trained to minimize the speech enhancement-related loss function at the MFMVDR filter output (see Section \ref{sec:settings}).

\subsection{A-Priori SNR}
\label{sec:a-priori snr}
Similarly to the approach described in the previous section, we propose to use a DNN $\f{\xi}$ to map noisy speech features to an a-priori SNR estimate $\hxi$, i.e.,
\begin{empheq}[box=\fbox]{align}
	\hxi = \f{\xi} \left( \log_{10} |Y_l| \right)
	\label{eq:a-priori snr}
\end{empheq}
Since $\xi_l \geq 0$, a softmax activation is used for $\f{\xi}$.
Similarly as for the correlation matrices, the DNN is trained to output an a-priori SNR estimate $\hxi$ such that the speech enhancement-related loss function at the MFMVDR filter output is minimized (see Section \ref{sec:settings}).

\section{Experimental Results}
In this section, the speech enhancement performance of the proposed deep MFMVDR filter is compared to several baseline deep learning-based speech enhancement algorithms (see Section \ref{sec:baselines}).
In Sections \ref{sec:dataset} - \ref{sec:settings} we discuss the used dataset, DNN architecture, and algorithm settings.
In Section \ref{sec:results} we present the results in terms of the perceptual evaluation of speech quality (PESQ)~\cite{itu-t_perceptual_2001} and short-time objective intelligibility (STOI)~\cite{taal_algorithm_2011} improvement.

\subsection{Baseline Algorithms}
\label{sec:baselines}
As baseline single-microphone speech enhancement algorithms, we consider three deep learning-based algorithms:
\begin{enumerate}
	\item \emph{Masking}: in order to investigate the possible benefit of multi-frame filtering, i.e., $N > 1$, we also consider the complex-valued gain/mask in \eqref{eq:masking}, i.e., $N = 1$:
	\begin{equation}
		W_{M,l} = \f{M} \left( \mathbf{y}_{c,l} \right) \in \mathbb{C}; \ \Re W_{\mathrm{M},l}, \Im W_{\mathrm{M},l} \in [-2,2],
	\end{equation}
	where the bounds for the real and imaginary parts are motivated by \cite{le_roux_phasebook_2019}.
	\item \emph{Direct filtering}: in order to investigate the possible benefit of using the MFMVDR filter structure in \eqref{eq:MFMVDR}, we also consider directly estimating the complex-valued coefficients of the $N$-dimensional multi-frame filter $\mathbf{w}_l$ in \eqref{eq:speech estimate vector}, similarly to \cite{mack_deep_2019}:
	\begin{equation}
		\mathbf{w}_{F,l} = \mbf{F} \left( \mathbf{y}_{c,l} \right) \in \mathbb{C}^N; \ \Re \mathbf{w}_{\mathrm{F},l}, \Im \mathbf{w}_{F,l} \in [-1,1],
	\end{equation}
	where the bounds for the real and imaginary parts are motivated by \cite{mack_deep_2019}.
	\item \emph{Conv-TasNet}~\cite{luo_conv-tasnet_2019}: instead of considering the STFT-domain as the transform-domain for speech enhancement, Conv-TasNet uses learnable transformations and applies a real-valued mask in the transform-domain.
	For a fair comparison with the other considered algorithms, we considered the causal version of Conv-TasNet~\cite{luo_conv-tasnet_2019}.
\end{enumerate}

\subsection{Dataset}
\label{sec:dataset}
All considered algorithms were trained and evaluated on the DNS Challenge dataset~\cite{reddy_interspeech_2020}.
In total, this dataset contains more than \unit[500]{h} of speech from 2150 speakers and \unit[180]{h} of noise from 150 different noise classes at a sampling frequency of \unit[16]{kHz}.
For training and validation, we randomly selected a subset of 45000 utterances of length \unit[4]{s}, with SNRs uniformly sampled from \unit[{[0, 20]}]{dB}.
Using a validation split of \unit[20]{$\%$}, this resulted in \unit[40]{h} for training and \unit[10]{h} for validation, respectively.
Evaluation was performed on the DNS Challenge synthetic test set without reverberation.
This test set is disjoint from the training and validation set and includes 20 speakers, 12 VoIP-relevant noise sources, and SNRs uniformly sampled from \unit[{[0, 25]}]{dB}, in total consisting of 150 utterances of length \unit[10]{s}.

\subsection{DNN Architecture}
\label{sec:dnn architecture}
As the DNN architecture for all estimators, we used temporal convolutional networks (TCNs)~\cite{bai_empirical_2018}\footnote{We used the implementation provided by the authors of \cite{luo_conv-tasnet_2019}, available at \href{https://github.com/naplab/Conv-TasNet}{https://github.com/naplab/Conv-TasNet}.}, which have been demonstrated to exhibit strong temporal and spectral modeling capabilities~\cite{luo_conv-tasnet_2019}.
Without performing extensive hyperparameter optimization, we fixed the hyperparameters of all TCN modules (except for the Conv-TasNet baseline, for which we used the hyperparameters proposed in \cite{luo_conv-tasnet_2019}) to 2 stacks of 4 layers each, with a kernel size of 3, resulting in a temporal receptive field of \unit[128]{ms}.
Aiming at a fair comparison, the number of hidden dimensions was varied to obtain a similar total number of trainable weights for all considered algorithms (cf. Table \ref{tab:hyperparams}).
Note that this hyperparameter was varied as opposed to the number of stacks/layers or the kernel size, since increasing the temporal receptive field might give an unfair advantage.
\begin{table}
	\begin{tabularx}{\linewidth}{l|l|l}
		algorithm & hidden dimension & trainable weights\\
		\toprule[2pt]
		masking & 226 & \unit[5.0]{M}\\
		direct filtering & 225 & \unit[5.1]{M}\\
		Conv-TasNet & 128 & \unit[5.0]{M}\\
		\midrule[1pt]
		deep MFMVDR & 128 &\unit[5.3]{M}
	\end{tabularx}
	\caption{TCN module hyperparameters.}
	\label{tab:hyperparams}
\end{table}

\subsection{Algorithm Settings}
\label{sec:settings}
In order to be able to exploit speech correlation, an STFT with high temporal resolution, i.e., a frame length of \unit[8]{ms} and a frame shift of \unit[2]{ms}, was employed for all STFT-based algorithms, similarly as in~\cite{huang_multi-frame_2012}.
A Hann window was used both as analysis and synthesis window.
The multi-frame algorithms (i.e., the proposed deep MFMVDR filter and direct filtering) use a filter length of $N=5$, such that speech correlation within \unit[16]{ms} can be exploited.
To improve the numerical stability of the matrix inversion in \eqref{eq:MFMVDR}, Tikhonov regularization with a regularization constant $\delta=10^{-3}$ was used~\cite{huang_multi-frame_2012,schasse_estimation_2014}.
Finally, a minimum gain of \unit[-17]{dB} was included in all algorithms except Conv-TasNet.

The TCNs were trained for 50 epochs using the Adam optimizer~\cite{kingma_adam:_2014} with an initial learning rate of $3 * 10^{-4}$.
The learning rate was halved after the validation loss did not decrease for 3 consecutive epochs, and training was stopped early in case the validation loss did not decrease for 10 consecutive epochs.
The gradient norms were clipped to 5, and the batch size was set to 6 to maximize the usage of graphics card memory.
As loss function, we used the negative scale-invariant signal-to-distortion ratio (SI-SDR)~\cite{roux_sdr_2019}, i.e.,
\begin{equation}
	\textnormal{SI-SDR} = 10 \log_{10} \left( \frac{|\alpha \tilde{\mathbf{x}}|^2}{|\alpha \tilde{\mathbf{x}} - \widehat{\tilde{\mathbf{x}}}|^2} \right), \quad
	\alpha = \frac{\widehat{\tilde{\mathbf{x}}}\tran \tilde{\mathbf{x}}}{||\tilde{\mathbf{x}}||^2},
	\label{eq:loss}
\end{equation}
where $\tilde{\mathbf{x}}$ and $\widehat{\tilde{\mathbf{x}}}$ denote the speech signal and the estimated speech signal in the time-domain.
All algorithms were implemented in PyTorch 1.6.0, and training and evaluation were performed on an NVIDIA GeForce\textsuperscript{\textregistered} RTX 2080 Ti graphics card.

\subsection{Results}
\label{sec:results}
For all considered algorithms, Table \ref{tab:results} shows the average improvement in terms of PESQ and STOI w.r.t. the noisy microphone signals using the speech signal as reference signal.
As can be observed, all considered algorithms yield a significant PESQ and STOI improvement, where the proposed deep MFMVDR filter outperforms all other algorithms.
A minor performance improvement can be observed between direct filtering ($N$~=~$5$) and masking ($N$~=~$1$), hinting at the potential of exploiting multiple frames.
A much larger improvement can be observed between deep MFMVDR filtering and direct filtering, showing that exploiting the MFMVDR filter structure and guiding the TCN training to estimate the required quantities (correlation matrices and a-priori SNR) instead of directly estimating the filter coefficients is advantageous.
Exemplary audio examples for all considered algorithms are available online\footnote{\scriptsize\href{https://uol.de/en/mediphysics-acoustics/sigproc/research/audio-demos}{https://uol.de/en/mediphysics-acoustics/sigproc/research/audio-demos}}.

As a measure for computational complexity, Table \ref{tab:results} also shows the average real-time factor (RTF) for all considered algorithms, defined as $\textnormal{RTF} = \textnormal{(processing time)} / \textnormal{(signal length})$, processed using 4 cores of an Intel\textsuperscript{\textregistered} Xeon\textsuperscript{\textregistered} CPU clocked at \unit[2.6]{GHz}.
Although the proposed deep MFMVDR filter results in a larger computational complexity than directly estimating the filter coefficients, its computational complexity is similar to that of Conv-TasNet.
A PyTorch implementation of the deep MFMVDR filter is available online\footnote{\scriptsize\href{https://uol.de/en/mediphysics-acoustics/sigproc/research/code-examples}{https://uol.de/en/mediphysics-acoustics/sigproc/research/code-examples}}.

\begin{table}
	\begin{tabularx}{\linewidth}{l|l|l|l}
		algorithm & $\Delta$PESQ / MOS & $\Delta$STOI & real-time factor\\
		\toprule[2pt]
		masking & 0.65 & 0.037 & \textbf{0.068}\\
		direct filtering & 0.67 & 0.038 & 0.070\\
		Conv-TasNet & 0.67 & 0.041 & 0.194\\
		\midrule[1pt]
		deep MFMVDR & \textbf{0.76} & \textbf{0.042} & 0.176
	\end{tabularx}
	\caption{PESQ and STOI improvement as well as real-time factors obtained on the DNS Challenge synthetic test set without reverberation, averaged over all utterances.}
	\label{tab:results}
\end{table}

\section{Conclusion}
In this paper we proposed a supervised learning-based approach to estimate the required parameters of an MFMVDR filter for single-microphone speech enhancement.
Because the MFMVDR filter requires accurate estimates of the noisy speech and noise correlation matrices as well as the speech IFC vector, we proposed to utilize the temporal and spectral modeling capabilities of TCNs for this estimation task.
The TCNs are trained to map the noisy speech STFT coefficients to the required parameters by minimizing the SI-SDR loss function at the output of the MFMVDR filter.
Experiments on the DNS Challenge dataset demonstrate the benefits of (i) using a multi-frame algorithm as compared to a single-frame algorithm, and (ii) guiding the TCN training by using the MFMVDR filter structure instead of directly estimating the filter coefficients.

%\section{REFERENCES}
% References should be produced using the bibtex program from suitable
% BiBTeX files (here: strings, refs, manuals). The IEEEbib.bst bibliography
% style file from IEEE produces unsorted bibliography list.
% -------------------------------------------------------------------------
\bibliographystyle{IEEEbib}
\bibliography{refs}

\end{document}

%% file: block_diagram_icassp2021.pdf_tex
%% Creator: Inkscape 1.0.1 (1.0.1+r73), www.inkscape.org
%% PDF/EPS/PS + LaTeX output extension by Johan Engelen, 2010
%% Accompanies image file 'block_diagram_icassp2021.pdf' (pdf, eps, ps)
%%
%% To include the image in your LaTeX document, write
%%   \input{<filename>.pdf_tex}
%%  instead of
%%   \includegraphics{<filename>.pdf}
%% To scale the image, write
%%   \def\svgwidth{<desired width>}
%%   \input{<filename>.pdf_tex}
%%  instead of
%%   \includegraphics[width=<desired width>]{<filename>.pdf}
%%
%% Images with a different path to the parent latex file can
%% be accessed with the `import' package (which may need to be
%% installed) using
%%   \usepackage{import}
%% in the preamble, and then including the image with
%%   \import{<path to file>}{<filename>.pdf_tex}
%% Alternatively, one can specify
%%   \graphicspath{{<path to file>/}}
%% 
%% For more information, please see info/svg-inkscape on CTAN:
%%   http://tug.ctan.org/tex-archive/info/svg-inkscape
%%
\begingroup%
  \makeatletter%
  \providecommand\color[2][]{%
    \errmessage{(Inkscape) Color is used for the text in Inkscape, but the package 'color.sty' is not loaded}%
    \renewcommand\color[2][]{}%
  }%
  \providecommand\transparent[1]{%
    \errmessage{(Inkscape) Transparency is used (non-zero) for the text in Inkscape, but the package 'transparent.sty' is not loaded}%
    \renewcommand\transparent[1]{}%
  }%
  \providecommand\rotatebox[2]{#2}%
  \newcommand*\fsize{\dimexpr\f@size pt\relax}%
  \newcommand*\lineheight[1]{\fontsize{\fsize}{#1\fsize}\selectfont}%
  \ifx\svgwidth\undefined%
    \setlength{\unitlength}{422.9270058bp}%
    \ifx\svgscale\undefined%
      \relax%
    \else%
      \setlength{\unitlength}{\unitlength * \real{\svgscale}}%
    \fi%
  \else%
    \setlength{\unitlength}{\svgwidth}%
  \fi%
  \global\let\svgwidth\undefined%
  \global\let\svgscale\undefined%
  \makeatother%
  \begin{picture}(1,0.20203211)%
    \lineheight{1}%
    \setlength\tabcolsep{0pt}%
    \put(0,0){\includegraphics[width=\unitlength,page=1]{block_diagram_icassp2021.pdf}}%
    \put(0.8819273,0.08432238){\color[rgb]{0,0,0}\makebox(0,0)[t]{\lineheight{1.25}\smash{\begin{tabular}[t]{c}loss \eqref{eq:loss}\end{tabular}}}}%
    \put(0,0){\includegraphics[width=\unitlength,page=2]{block_diagram_icassp2021.pdf}}%
    \put(0.68271096,0.02413722){\color[rgb]{0,0,0}\makebox(0,0)[lt]{\lineheight{1.25}\smash{\begin{tabular}[t]{l}:trainable\end{tabular}}}}%
    \put(0,0){\includegraphics[width=\unitlength,page=3]{block_diagram_icassp2021.pdf}}%
    \put(0.78178486,0.0241372){\color[rgb]{0,0,0}\makebox(0,0)[lt]{\lineheight{1.25}\smash{\begin{tabular}[t]{l}:non-trainable\end{tabular}}}}%
    \put(0,0){\includegraphics[width=\unitlength,page=4]{block_diagram_icassp2021.pdf}}%
    \put(0.60232619,0.09655921){\color[rgb]{0,0,0}\makebox(0,0)[t]{\lineheight{1.25}\smash{\begin{tabular}[t]{c}MFMVDR\\filter \eqref{eq:MFMVDR}\end{tabular}}}}%
    \put(0.08564073,0.0849539){\color[rgb]{0,0,0}\makebox(0,0)[t]{\lineheight{1.25}\smash{\begin{tabular}[t]{c}+\end{tabular}}}}%
    \put(0,0){\includegraphics[width=\unitlength,page=5]{block_diagram_icassp2021.pdf}}%
    \put(0.32147359,0.10809544){\color[rgb]{0,0,0}\makebox(0,0)[t]{\lineheight{1.25}\smash{\begin{tabular}[t]{c}$\f{\xi}$\\a-priori\\SNR \eqref{eq:a-priori snr}\end{tabular}}}}%
    \put(0,0){\includegraphics[width=\unitlength,page=6]{block_diagram_icassp2021.pdf}}%
    \put(0.74233915,0.08444589){\color[rgb]{0,0,0}\makebox(0,0)[t]{\lineheight{1.25}\smash{\begin{tabular}[t]{c}filter \eqref{eq:speech estimate vector}\end{tabular}}}}%
    \put(0,0){\includegraphics[width=\unitlength,page=7]{block_diagram_icassp2021.pdf}}%
    \put(0.18159259,0.16803614){\color[rgb]{0,0,0}\makebox(0,0)[t]{\lineheight{1.25}\smash{\begin{tabular}[t]{c}\footnotesize correlation\\\footnotesize matrices \eqref{eq:correlation matrix states}, \eqref{eq:correlation matrices}\end{tabular}}}}%
    \put(0,0){\includegraphics[width=\unitlength,page=8]{block_diagram_icassp2021.pdf}}%
    \put(0.18092182,0.10504359){\makebox(0,0)[t]{\lineheight{1.25}\smash{\begin{tabular}[t]{c}$\mbf{y}$\end{tabular}}}}%
    \put(0,0){\includegraphics[width=\unitlength,page=9]{block_diagram_icassp2021.pdf}}%
    \put(0.18091288,0.06306605){\makebox(0,0)[t]{\lineheight{1.25}\smash{\begin{tabular}[t]{c}$\mbf{n}$\end{tabular}}}}%
    \put(0.37684333,0.00459272){\makebox(0,0)[t]{\lineheight{1.25}\smash{\begin{tabular}[t]{c}$\bhPhi{y}, \ \bhPhi{n}$\end{tabular}}}}%
    \put(0.39130765,0.11069834){\makebox(0,0)[t]{\lineheight{1.25}\smash{\begin{tabular}[t]{c}$\hxi$\end{tabular}}}}%
    \put(0.53144489,0.11218422){\makebox(0,0)[t]{\lineheight{1.25}\smash{\begin{tabular}[t]{c}$\bhgamma{x}$\end{tabular}}}}%
    \put(0.67158254,0.11218422){\makebox(0,0)[t]{\lineheight{1.25}\smash{\begin{tabular}[t]{c}$\mathbf{w}_{l}$\end{tabular}}}}%
    \put(0.81171947,0.11218422){\makebox(0,0)[t]{\lineheight{1.25}\smash{\begin{tabular}[t]{c}$\widehat{X}_{l}$\end{tabular}}}}%
    \put(0,0){\includegraphics[width=\unitlength,page=10]{block_diagram_icassp2021.pdf}}%
    \put(0.46212383,0.09626828){\color[rgb]{0,0,0}\makebox(0,0)[t]{\lineheight{1.25}\smash{\begin{tabular}[t]{c}speech IFC\\vector \eqref{eq:speech IFC vector estimate}\end{tabular}}}}%
  \end{picture}%
\endgroup%

%% file: main.bbl
\begin{thebibliography}{10}

\bibitem{vary_digital_2006}
P.~Vary and R.~Martin,
\newblock {\em Digital speech transmission: enhancement, coding and error
  concealment},
\newblock John Wiley, Chichester, England; Hoboken, NJ, 2006.

\bibitem{hendriks_dft-domain_2013}
R.~C. Hendriks, T.~Gerkmann, and J.~Jensen,
\newblock {\em {DFT}-{Domain} {Based} {Single}-{Microphone} {Noise} {Reduction}
  for {Speech} {Enhancement}: {A} {Survey} of the {State} of the {Art}},
\newblock Morgan \& Claypool Publishers, 2013.

\bibitem{benesty_speech_2011}
J.~Benesty, J.~Chen, and E.~A.~P. Habets,
\newblock {\em Speech {Enhancement} in the {STFT} {Domain}},
\newblock Springer Science \& Business Media, 2011.

\bibitem{doclo_multichannel_2015}
S.~Doclo, W.~Kellermann, S.~Makino, and S.~E. Nordholm,
\newblock ``Multichannel signal enhancement algorithms for assisted listening
  devices: {Exploiting} spatial diversity using multiple microphones,''
\newblock {\em IEEE Signal Processing Magazine}, vol. 32, no. 2, pp. 18--30,
  Mar. 2015.

\bibitem{vincent_audio_2018}
E.~Vincent, T.~Virtanen, and S.~Gannot,
\newblock {\em Audio {Source} {Separation} and {Speech} {Enhancement}},
\newblock John Wiley \& Sons, 2018.

\bibitem{ephraim_speech_1984}
Y.~Ephraim and D.~Malah,
\newblock ``Speech enhancement using a minimum mean-square error short-time
  spectral amplitude estimator,''
\newblock {\em IEEE Trans. Acoustics, Speech and Signal Processing}, vol. 32,
  no. 6, pp. 1109--1121, Dec. 1984.

\bibitem{gerkmann_phase_2015}
T.~Gerkmann, M.~Krawczyk-Becker, and J.~Le Roux,
\newblock ``Phase {Processing} for {Single}-{Channel} {Speech} {Enhancement}:
  {History} and recent advances,''
\newblock {\em IEEE Signal Processing Magazine}, vol. 32, no. 2, pp. 55--66,
  Mar. 2015.

\bibitem{xu_regression_2015}
Y.~Xu, J.~Du, L.~Dai, and C.~Lee,
\newblock ``A {Regression} {Approach} to {Speech} {Enhancement} {Based} on
  {Deep} {Neural} {Networks},''
\newblock {\em IEEE/ACM Trans. Audio, Speech and Language Processing}, vol. 23,
  no. 1, pp. 7--19, Jan. 2015.

\bibitem{wang_training_2014}
Y.~Wang, A.~Narayanan, and D.~L. Wang,
\newblock ``On {Training} {Targets} for {Supervised} {Speech} {Separation},''
\newblock {\em IEEE/ACM Trans. Audio, Speech, and Language Processing}, vol.
  22, no. 12, pp. 1849--1858, Dec. 2014.

\bibitem{kolbaek_speech_2017}
M.~Kolbæk, Z.~Tan, and J.~Jensen,
\newblock ``Speech {Intelligibility} {Potential} of {General} and {Specialized}
  {Deep} {Neural} {Network} {Based} {Speech} {Enhancement} {Systems},''
\newblock {\em IEEE/ACM Trans. Audio, Speech, and Language Processing}, vol.
  25, no. 1, pp. 153--167, Jan. 2017.

\bibitem{luo_conv-tasnet_2019}
Y.~Luo and N.~Mesgarani,
\newblock ``Conv-{TasNet}: {Surpassing} {Ideal} {Time}–{Frequency}
  {Magnitude} {Masking} for {Speech} {Separation},''
\newblock {\em IEEE/ACM Trans. Audio, Speech, and Language Processing}, vol.
  27, no. 8, pp. 1256--1266, Aug. 2019.

\bibitem{tan_gated_2019}
K.~Tan, J.~Chen, and D.~Wang,
\newblock ``Gated {Residual} {Networks} {With} {Dilated} {Convolutions} for
  {Monaural} {Speech} {Enhancement},''
\newblock {\em IEEE/ACM Trans. Audio, Speech, and Language Processing}, vol.
  27, no. 1, pp. 189--198, Jan. 2019.

\bibitem{huang_multi-frame_2012}
Y.~A. Huang and J.~Benesty,
\newblock ``A {Multi}-{Frame} {Approach} to the {Frequency}-{Domain}
  {Single}-{Channel} {Noise} {Reduction} {Problem},''
\newblock {\em IEEE Trans. Audio, Speech, and Language Processing}, vol. 20,
  no. 4, pp. 1256--1269, May 2012.

\bibitem{schasse_estimation_2014}
A.~Schasse and R.~Martin,
\newblock ``Estimation of {Subband} {Speech} {Correlations} for {Noise}
  {Reduction} via {MVDR} {Processing},''
\newblock {\em IEEE/ACM Trans. Audio, Speech, and Language Processing}, vol.
  22, no. 9, pp. 1355--1365, Sept. 2014.

\bibitem{fischer_sensitivity_2017}
D.~Fischer and S.~Doclo,
\newblock ``Sensitivity analysis of the multi-frame {MVDR} filter for
  single-microphone speech enhancement,''
\newblock in {\em Proc. {European} {Signal} {Processing} {Conference}
  ({EUSIPCO})}, Kos, Greece, Aug. 2017, pp. 603--607.

\bibitem{fischer_subspace-based_2020}
D.~Fischer and S.~Doclo,
\newblock ``Subspace-{Based} {Speech} {Correlation} {Vector} {Estimation} for
  {Single}-{Microphone} {Multi}-{Frame} {MVDR} {Filtering},''
\newblock in {\em Proc. {IEEE} {International} {Conference} on {Acoustics},
  {Speech} and {Signal} {Processing} ({ICASSP})}, Barcelona, Spain, May 2020,
  pp. 856--860.

\bibitem{mack_deep_2019}
W.~Mack and E.~A.~P. Habets,
\newblock ``Deep {Filtering}: {Signal} {Extraction} and {Reconstruction}
  {Using} {Complex} {Time}-{Frequency} {Filters},''
\newblock {\em IEEE Signal Processing Letters}, vol. 27, Nov. 2019.

\bibitem{xu_neural_2020}
Y.~Xu, M.~Yu, S.-X. Zhang, L.~Chen, C.~Weng, J.~Liu, and D.~Yu,
\newblock ``Neural {Spatio}-{Temporal} {Beamformer} for {Target} {Speech}
  {Separation},''
\newblock {\em arXiv:2005.03889 [cs, eess]}, May 2020.

\bibitem{tammen_dnn-based_2020}
M.~Tammen, D.~Fischer, B.~T. Meyer, and S.~Doclo,
\newblock ``{DNN}-{Based} {Speech} {Presence} {Probability} {Estimation} for
  {Multi}-{Frame} {Single}-{Microphone} {Speech} {Enhancement},''
\newblock in {\em Proc. {IEEE} {International} {Conference} on {Acoustics},
  {Speech} and {Signal} {Processing} ({ICASSP})}, Barcelona, Spain, May 2020,
  pp. 191--195.

\bibitem{bai_empirical_2018}
S.~Bai, J.~Z. Kolter, and V.~Koltun,
\newblock ``An {Empirical} {Evaluation} of {Generic} {Convolutional} and
  {Recurrent} {Networks} for {Sequence} {Modeling},''
\newblock {\em arXiv:1803.01271 [cs]}, Mar. 2018.

\bibitem{roux_sdr_2019}
J.~Le Roux, S.~Wisdom, H.~Erdogan, and J.~R. Hershey,
\newblock ``{SDR} – {Half}-baked or {Well} {Done}?,''
\newblock in {\em Proc. {IEEE} {International} {Conference} on {Acoustics},
  {Speech} and {Signal} {Processing} ({ICASSP})}, Brighton, UK, May 2019, pp.
  626--630.

\bibitem{reddy_interspeech_2020}
C.~K.~A. Reddy, V.~Gopal, R.~Cutler, E.~Beyrami, R.~Cheng, H.~Dubey,
  S.~Matusevych, R.~Aichner, A.~Aazami, S.~Braun, P.~Rana, S.~Srinivasan, and
  J.~Gehrke,
\newblock ``The {INTERSPEECH} 2020 {Deep} {Noise} {Suppression} {Challenge}:
  {Datasets}, {Subjective} {Testing} {Framework}, and {Challenge} {Results},''
\newblock {\em arXiv:2005.13981 [cs, eess]}, May 2020.

\bibitem{itu-t_perceptual_2001}
ITU-T,
\newblock ``Perceptual evaluation of speech quality ({PESQ}), an objective
  method for end-to-end speech quality assessment of narrowband telephone
  networks and speech codecs {P}.862,''
\newblock Tech. {R}ep., International Telecommunications Union (ITU-T)
  Recommendation, Feb. 2001.

\bibitem{taal_algorithm_2011}
C.~H. Taal, R.~C. Hendriks, R.~Heusdens, and J.~Jensen,
\newblock ``An {Algorithm} for {Intelligibility} {Prediction} of
  {Time}–{Frequency} {Weighted} {Noisy} {Speech},''
\newblock {\em IEEE Trans. Audio, Speech, and Language Processing}, vol. 19,
  no. 7, pp. 2125--2136, Sept. 2011.

\bibitem{le_roux_phasebook_2019}
J.~Le~Roux, G.~Wichern, S.~Watanabe, A.~Sarroff, and J.~R. Hershey,
\newblock ``Phasebook and {Friends}: {Leveraging} {Discrete} {Representations}
  for {Source} {Separation},''
\newblock {\em IEEE Journal of Selected Topics in Signal Processing}, vol. 13,
  no. 2, pp. 370--382, May 2019.

\bibitem{kingma_adam:_2014}
D.~P. Kingma and J.~Ba,
\newblock ``Adam: {A} {Method} for {Stochastic} {Optimization},''
\newblock {\em arXiv:1412.6980 [cs]}, Dec. 2014.

\end{thebibliography}
